\documentclass[%
 reprint,
 amsmath,amssymb,
 aps,
]{revtex4-2}

\usepackage{graphicx}
\usepackage{dcolumn}
\usepackage{bm}
\usepackage{diagbox}
\usepackage{makecell}

\begin{document}

\preprint{APS/123-QED}

\title{Equal-Spin local and nonlocal Andreev reflection in altermagnet/superconductor/altermagnet junction}

\author{Kai-Yi Lyu\textsuperscript{1}}

\author{Xing Wang\textsuperscript{2,3}}

\author{Yu-Xian Li\textsuperscript{1}}

\email{Contact author: yxli@mail.hebtu.edu.cn}

\address{$^{1}$College of Physics and Hebei Advanced Thin Films Laboratory, Hebei Normal University, Shijiazhuang, Hebei 050024, China\\$^{2}$Department of Physics, Tsinghua University, Beijing 100084, People's Republic of China\\$^{3}$State Key Laboratory of Low-Dimensional Quantum Physics, Tsinghua University, Beijing 100084, People's Republic of China}

\date{\today}

\begin{abstract}
We investigated the equal-spin Andreev reflection phenomenon in an altermagnet/superconductor/altermagnet (AM/SC/AM) junction, focusing on the roles of the crystal orientation angle and the Néel vector. The crystal orientation angle in the AM effectively determines the spin-dependent transport channels for both electrons and holes, leading to orientation-dependent local and crossed Andreev reflection in the AM/SC/AM junction. We studied how and under what conditions this equal-spin Andreev reflection occurs. Rotating the Néel vector in the AM substantially increases spin-flip channels for both electrons and holes, giving rise to equal-spin Andreev reflection. When the Néel vector $\theta=\pi/2$, the spin-dependent transmission channels for crystal orientations $\alpha=0$ and $\alpha=\pi/2$ are completely the same, the Andreev reflection coefficients remain different, reflecting the distinct roles of the Néel vector in spin space and the crystal orientation in real space. By tuning the crystal orientation angle and the Néel vector, orientation-dependent conventional and equal-spin Andreev reflection can be realized. This study reveals that Néel vector rotation is the direct cause of equal-spin Andreev reflection in AM/SC/AM junctions and provides a viable pathway for realizing spin-dependent superconducting transport phenomena in AM heterojunctions.

\end{abstract}
\maketitle

\section{\label{sec:level1}INTRODUCTION}

In recent years, altermagnetism has attracted significant attention in condensed matter physics \cite{1,2,3,4,5,6}. The altermagnet (AM) exhibits zero net magnetization due to the staggered spatial arrangement of magnetic moments, similar to antiferromagnets, while simultaneously hosting an alternating spin polarization in momentum space, analogous to ferromagnets, which breaks time-reversal symmetry. Consequently, altermagnetism is regarded as a third magnetic phase beyond ferromagnetism and antiferromagnetism \cite{7,8}. A number of materials have been proposed to realize altermagnetic order \cite{9,10}, among which RuO$_2$, MnSi$_3$, MnF$_2$, and MnO$_2$ are the most extensively studied d-wave altermagnets. Moreover, the unique band structure of AM gives rise to a variety of novel physical phenomena, including giant magnetoresistance and tunnel magnetoresistance \cite{12,15}, anomalous Hall effects \cite{11,13,16}, and efficient spin-to-charge conversion \cite{14}. Owing to the absence of macroscopic net magnetization, AM avoid stray magnetic fields that commonly appear in ferromagnetic systems. This makes AM an ideal magnetic material in the design and construction of heterojunctions consisting of AM and superconductor (SC). Recent studies have investigated Andreev reflection \cite{17,18,46,18a}, dc Josephson effects \cite{5,19,20,20a}, and Majorana corner modes \cite{21,22} in AM/SC hybrid systems, revealing crystal orientation–dependent transport characteristics that originate from the anisotropic spin polarization inherent to altermagnets.

Andreev reflection (AR) is an electron–hole conversion process occurring at the interface between a superconductor and a normal conductor, in which an incident electron is converted into a hole while a Cooper pair is simultaneously formed inside the superconductor \cite{23}. When the incident electron and the reflected hole reside in the same lead, the process is referred to as local Andreev reflection (LAR) \cite{24,25,26}, whereas crossed Andreev reflection (CAR) is a nonlocal process in which the electron and hole are located in two spatially separated leads \cite{27,28,29,29a}. CAR enables the spatial separation of Cooper pairs, giving rise to nonlocally entangled electrons that are essential resources for quantum communication and quantum computing \cite{30,31,32}. In conventional AR, the incident electron and the Andreev-reflected hole belong to opposite spin bands, while in equal-spin AR the two quasiparticles occupy the same spin band, reflecting a spin-parity-preserving pairing process. Various mechanisms for generating equal-spin AR have been extensively explored, including ferromagnet/superconductor interfaces with inhomogeneous exchange fields in real space \cite{36,37,38,39,40,42} and systems with spin–orbit coupling in momentum space \cite{43,44}. When AM is in a strong phase, perfect CAR can be achieved \cite{45}. More recently, it has been demonstrated that equal-spin AR can also be induced by modulating the Néel vector in altermagnetic systems \cite{46,47}. Furthermore, when the applied bias voltage lies within the superconducting gap, charge transport in hybrid junctions is dominated by AR processes \cite{33,34,35}.

Inspired by the above studies, we investigate the combined effects of the crystal orientation angle and the Néel vector on AR in hybrid altermagnetic systems. In this work, we employ a tight-binding model together with the Landauer–Büttiker formalism and the non-equilibrium Green’s function method to study AR in an altermagnet/superconductor/altermagnet(AM/SC/AM) junction. Our results demonstrate that equal-spin AR emerges only when the Néel vector is rotated, highlighting the essential role of spin-flip induced by Néel vector modulation. By tuning the crystal orientation angle, AR can be selectively dominated by different types of incident electrons. Moreover, the combined effect of the Néel vector and the crystal orientation angle enables orientation-dependent equal-spin AR dominated by distinct incident electron species.

The remainder of this paper is organized as follows. In Sec. \ref{SC2}, we introduce the Hamiltonian and band structure of the AM/SC/AM junction, along with the formulas and methods used to calculate AR. In Sec. \ref{SC3}, we present numerical results and discussions on how the crystal orientation angle and the Néel vector affect the spin transmission channels in AMs and the AR coefficients in AM/SC/AM junctions. Finally, a summary is given in Sec. \ref{SC4}.

\begin{figure}[htbp]
\includegraphics[width=0.48\textwidth]{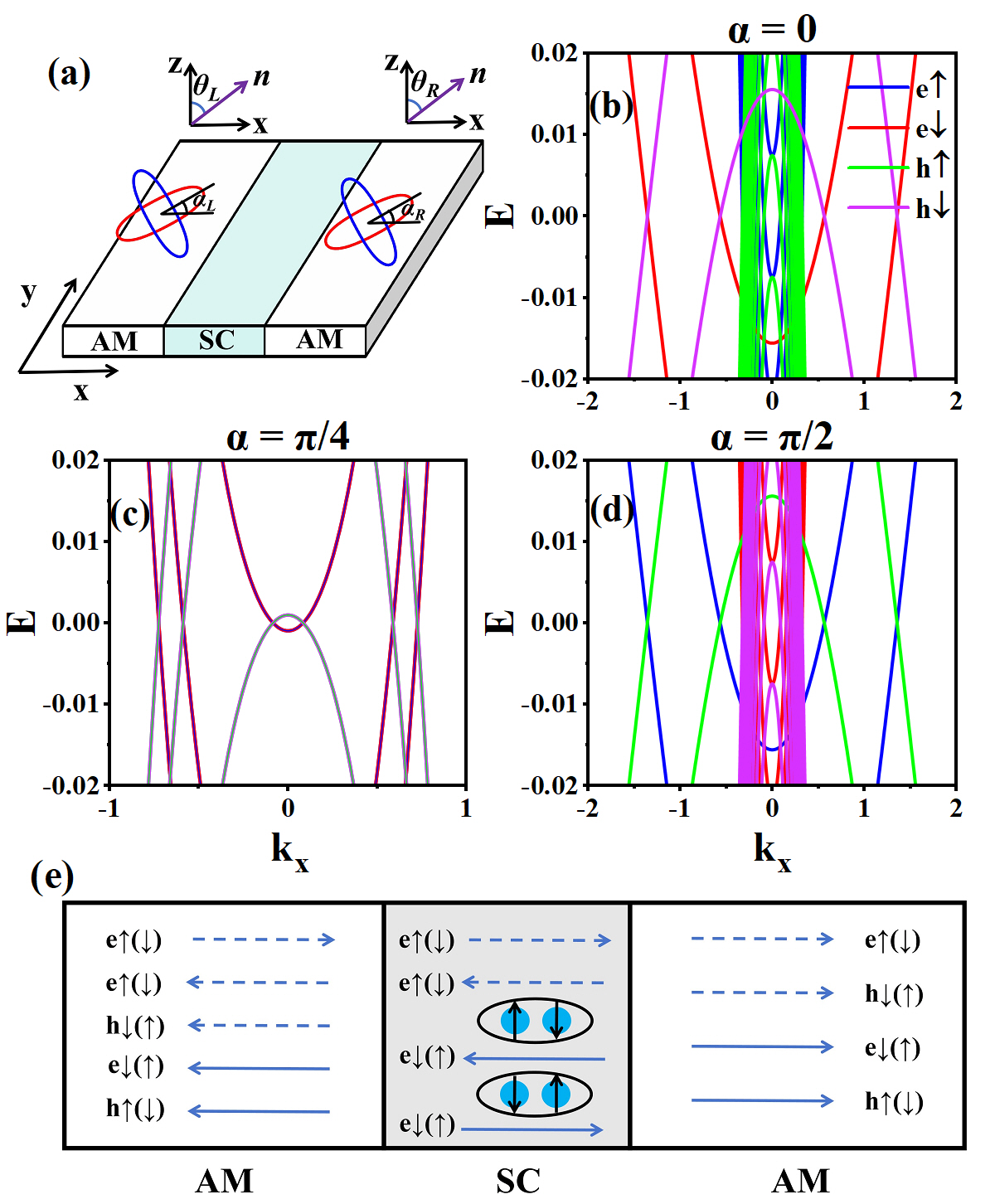}
\caption{\label{fig1}(a) Schematic of the AM/SC/AM junction. The central region is an s-wave superconducting region. The two ellipses represent the Fermi surfaces of electrons with opposite spins in the AM region: the blue (red) ellipse corresponds to spin-up (spin-down) electrons. The orientation angle of the AM is defined by $\alpha$ relative to the $x$ axis. The Néel vector $\bm{n}$ lies in the $x–z$ plane and forms an angle $\theta_{L(R)}$ with the $z$ axis. Band structures of the AM at $\theta_{L(R)}=0$ for different orientation angles $\alpha$: (b) $\alpha_{L(R)} = 0$, (c) $\alpha_{L(R)} = \pi/4$, and (d) $\alpha_{L(R)} = \pi/4$. Blue and red lines denote spin-up and spin-down electron bands, while green and purple lines represent spin-up and spin-down hole bands, respectively. (e) All scattering processes in the system. Dashed (solid) lines denote non–spin-flip (spin-flip) processes.}
\end{figure}

\section{MODEL AND METHODS}\label{SC2}

We consider an AM/SC/AM junction based on $d$-wave altermagnets in the $x$–$y$ plane, as schematically illustrated in Fig.~\ref{fig1}(a). The left and right regions consist of semi-infinite AM leads, while the central region is an $s$-wave superconducting region with length $L$ and width $W$. The two ellipses depict the Fermi surfaces associated with electrons of opposite spins in the AM, with the blue (red) ellipse corresponding to spin-up (spin-down) electrons. The AM crystallographic orientation is specified by the angle $\alpha$ with respect to the $x$ axis. The Néel vector $\bm{n}$ lies in the $x$–$z$ plane and forms an angle $\theta$ with the $z$ axis.

The system is described within the Bogoliubov–de Gennes (BdG) formalism in the Nambu basis $\Psi(\mathbf{k}) = (c_{\mathbf{k}\uparrow}, c_{\mathbf{k}\downarrow}, c^\dagger_{-\mathbf{k}\uparrow}, c^\dagger_{-\mathbf{k}\downarrow})^T$. This formalism enables the simultaneous treatment of superconducting pairing and spin-dependent transport in the altermagnetic system. The Hamiltonian for the systemt can be written as \cite{21}
\begin{eqnarray}\label{Eq1}
{H(\mathbf{k})=H_0(\mathbf{k})+H_J(\mathbf{k})+H_\Delta(\mathbf{k}),}
\end{eqnarray}
where 
\begin{eqnarray}
H_0(\mathbf{k}) &=& \left[t\left(2-\cos k_x-\cos k_y\right)-\mu\right]\tau_z \otimes \sigma_0, \nonumber\\
H_J(\mathbf{k}) &=& t_J\sin2\alpha\sin k_x\sin k_y \tau_z\otimes(\sigma_{i}\cdot\bm{n}) \nonumber \\
&& + \; t_J\cos2\alpha\left(\cos k_x-\cos k_y\right)\tau_z\otimes(\sigma_{i}\cdot\bm{n}), \nonumber\\
H_\Delta(\mathbf{k}) &=& \Delta_0 \tau_y \otimes \sigma_y.
\label{Eq2}
\end{eqnarray}
Here, $\tau_i$ and $\sigma_i$ ($i=x,y,z$ and $0$) represent the Pauli matrices and identity matrices acting on the particle-hole space and spin space, respectively. The terms $H_0(\mathbf{k})$, $H_J(\mathbf{k})$, and $H_\Delta(\mathbf{k})$ describe the kinetic energy of particles, the d-wave altermagnetism, and the superconductivity, respectively. The factor $\tau_z$ indicates that $H_0(\mathbf{k})$ and $H_J(\mathbf{k})$ act separately on the electron and hole sectors in particle-hole space, while $H_\Delta(\mathbf{k})$ couples the electron and hole sectors through the superconducting pairing potential. This $4\times4$ BdG Hamiltonian provides a unified description of altermagnetism and superconductivity in the AM/SC/AM junction.

For simplicity, we assume idealized high-transparency AM/SC interfaces without additional interface barriers or disorder. Such an assumption is reasonable since advances in experimental nanofabrication techniques have made it possible to realize atomically sharp and clean interfaces \cite{a1,a2}.

The left (right) AM Hamiltonian is given by $H_{L(R)} = \sum_{\mathbf{k}} \psi_{L(R)}^{\dagger}(\mathbf{k}) H_{L(R)}(\mathbf{k}) \psi_{L(R)}(\mathbf{k})$, with $H_{L(R)}(\mathbf{k})=H_0(\mathbf{k})+H_J(\mathbf{k})$ and $\psi_{L(R)}(\mathbf{k}) = \left(c_{L(R)\mathbf{k}\uparrow}, \ c_{L(R)\mathbf{k}\downarrow}, \ c_{L(R)-\mathbf{k}\uparrow}^{\dagger}, \ c_{L(R)-\mathbf{k}\downarrow}^{\dagger} \right)^{T}$. The Hamiltonian of the central superconducting region is given by $H_c(\mathbf{k})=H_0(\mathbf{k})+H_\Delta(\mathbf{k})$. Here, $\mathbf{k} = (k_x, k_y)$ is the wave vector, $t$ represents the hopping energy, and $\mu$ is the chemical potential. The parameters $t_j$ and $\Delta_0$ characterize  the magnitudes of altermagnetism and the superconducting gap. We have used $\tau_{i}$ and $\sigma_{i}$($i = x, y, z$ and $0$) to denote the three components of the Pauli matrices and the identity matrices in particle-hole space and spin space, respectively. The Néel vector, represented by the unit vector $\bm{n} = (\sin\theta \cos\phi,\ \sin\theta \sin\phi,\ \cos\theta)$, can be rotated experimentally \cite{6,48,50,51,52,53}. When $\phi = 0$ and $\theta = \pi/2$, the term $(\sigma_i \cdot \bm{n}) = \sigma_x$, implying that rotating the Néel vector by $\pi/2$ induces spin mixing between the spin-up and spin-down channels.

Next, we discretize the Hamiltonian $H$ of the system on a two-dimensional square lattice. By setting the lattice constant $a = 1$ and performing a Fourier transformation along the $x$ and $y$ directions, the discretized tight-binding Hamiltonian can be written as
\begin{eqnarray}
H &=& \sum_{\mathbf{i}}\left[\psi_{\mathbf{i}}^{\dagger}T_{0}\psi_{\mathbf{i}}+\psi_{\mathbf{i}}^{\dagger}T_{x}\psi_{\mathbf{i}+\hat{\mathbf{x}}}+\psi_{\mathbf{i}}^{\dagger}T_{y}\psi_{\mathbf{i}+\hat{\mathbf{y}}}\right. \nonumber \\
&& + \; \left. \psi_{\mathbf{i}}^{\dagger}T_{xy}\psi_{\mathbf{i}+\hat{\mathbf{x}}+\hat{\mathbf{y}}}+\psi_{\mathbf{i}}^{\dagger}T_{x\bar{y}}\psi_{\mathbf{i}+\hat{\mathbf{x}}-\hat{\mathbf{y}}}+\text{H.c.}\right].
\label{Eq3}
\end{eqnarray}
where $\psi_{\mathbf{i}} = \left(\psi_{{i}\uparrow}, \psi_{{i}\downarrow}, \psi_{{i}\uparrow}^{\dagger}, \psi_{{i}\downarrow}^{\dagger} \right)^{T}$ is the operator at the discrete site $\mathbf{i}$ in the center region, $\hat{\mathbf{x}}$ and $\hat{\mathbf{y}}$ represent unit vectors in the $x$ and $y$ directions, respectively. The corresponding on-site matrix, nearest-neighbor hopping matrix, and next-nearest-neighbor hopping matrix are respectively
\begin{eqnarray}
T_{0} &=& (2t - \mu) \tau_{z} \otimes \sigma_{0} + \Delta_{0} \tau_{y} \otimes \sigma_{y}, \nonumber \\
T_{x} &=& -\frac{t}{2}\tau_{z} \otimes \sigma_{0} 
        + \frac{t_{J}}{2}\cos 2\alpha \tau_{z} \otimes (\sigma_{i}\cdot\bm{n}), \nonumber \\
T_{y} &=& -\frac{t}{2}\tau_{z} \otimes \sigma_{0} 
        - \frac{t_{J}}{2}\cos 2\alpha \tau_{z} \otimes (\sigma_{i}\cdot\bm{n}), \nonumber \\
T_{xy} &=& \frac{t_{J}}{4}\sin 2\alpha \tau_{z} \otimes (\sigma_{i}\cdot\bm{n})
         = -T_{x\bar{y}}.
\label{Eq4}
\end{eqnarray}

In this work, we set $\phi = 0$, $t=1$, $t_J=0.9$, $\mu=0.1$, and $\Delta_{0}=0.01$, and present the band structures of the altermagnetic system near incident energy $E = 0$ for different orientations $\alpha$, with the Néel vector angle fixed at $\theta = 0$, as shown in Figs.~\ref{fig1}(b)-(d) for $\alpha = 0$, $\pi/4$, and $\pi/2$, respectively. Blue and red lines denote the spin-up and spin-down electron bands, while green and purple lines represent the spin-up and spin-down hole bands. The band structures demonstrate that the spin-up and spin-down bands overlap at $\alpha = \pi/4$, whereas the spin-up and spin-down bands are inverted at $\alpha = \pi/2$ relative to the case $\alpha = 0$, consistent with the evolution of the elliptical Fermi surfaces.

Based on the Landauer–Büttiker formula combined with the non-equilibrium Green’s function method, the local Andreev reflection coefficient ($T_{\mathrm{LAR}}$) and the crossed Andreev reflection coefficient ($T_{\mathrm{CAR}}$) for the AM/SC/AM junction can be obtained as \cite{18a,35} $T_{\mathrm{LAR}} = \mathrm{Tr}[ \Gamma_{ee}^{L} G^{r} \Gamma_{hh}^{L} G^{a}]$, $T_{\mathrm{CAR}} = \mathrm{Tr}\left[ \Gamma_{ee}^{L} G^{r} \Gamma_{hh}^{R} G^{a} \right]$, where ($E$) is the incident energy and $\Gamma_{p} = i\left(\Sigma_{p}^{r} - \Sigma_{p}^{a}\right)$ denotes the linewidth function of $p$ lead (($p = L, R$) for the left and right leads, respectively). The retarded Green’s function of the central scattering region is given by $G^{r}(E) = \left[ E - H_c - \Sigma_L^{r}(E) - \Sigma_R^{r}(E) \right]^{-1}$, where ($H_c$) is the Hamiltonian of the central region. The retarded (advanced) self-energy ($\Sigma_p^{r(a)}$) arises from the coupling between $p$ lead and the central region and can be expressed as $\Sigma_{p}^{r} = \Sigma_{p}^{a\dagger} = H_{cp} g_{p}^{r}H_{pc}$,
where $H_{cp(pc)}$ is the coupling matrix between $p$ lead and the central region, and $g_{p}^{r}$ is the surface Green’s function of the semi-infinite lead, which can be obtained using an iterative method \cite{54,55}.

In Fig.~\ref{fig1}(e), we illustrate all possible scattering processes in the AM/SC/AM junction.
The dashed and solid lines denote scattering processes without and with spin-flip, respectively, induced by the rotation of the Néel vector.
Taking an incident spin-up electron as an example, the electron can be directly reflected as a spin-up electron in the left lead, or undergo spin-flip and be reflected as a spin-down electron.
It can also be transmitted into the right lead while preserving its spin direction, or be transmitted as a spin-down electron if spin-flip occurs in the left lead.
Furthermore, when the Néel vector in the right lead is rotated, a transmitted spin-up electron may experience an additional spin-flip process and be converted into a spin-down electron.

In the center superconducting region, part of the incident spin-up electrons at the SC/AM interface can undergo spin-flip due to the effect from the right AM lead.
Spin-down electrons can form Cooper pairs with spin-up electrons inside the superconductor, resulting in equal-spin LAR.
Electrons that undergo spin-flip in the left lead can also participate in pairing processes in the superconducting region, contributing simultaneously to equal-spin LAR and equal-spin CAR.
By contrast, spin-up electrons without spin-flip form Cooper pairs with spin-down electrons, leading to conventional LAR and CAR, in which spin-down holes are reflected into the left or right leads.
If these spin-down holes further experience spin-flip, part of the conventional AR processes are converted into equal-spin LAR and equal-spin CAR.

\begin{figure}[htbp]
\includegraphics[width=0.48\textwidth]{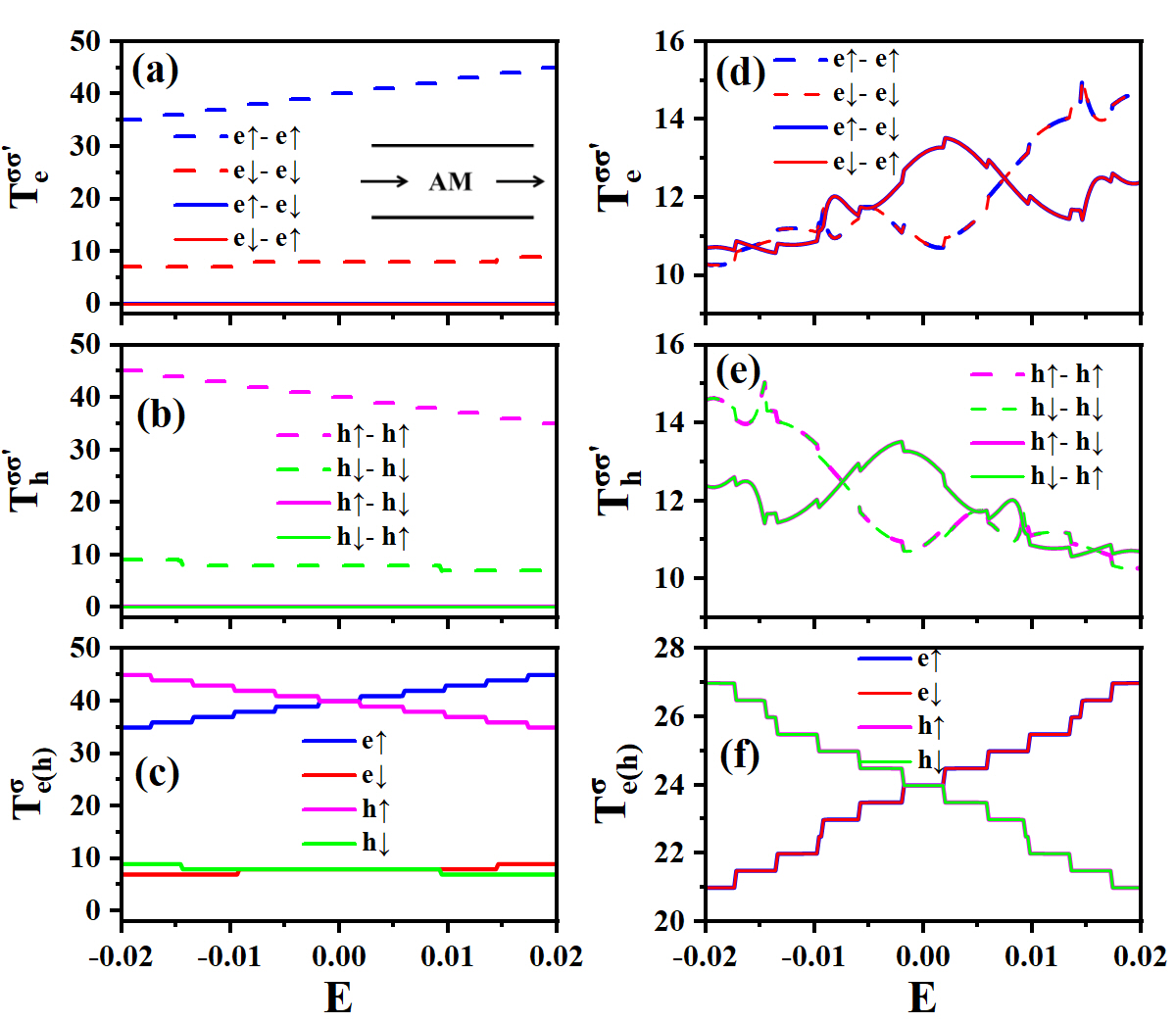}
\caption{\label{fig2} Transmission channels in the AM nanoribbon at a fixed orientation angle $\alpha = 0$. The inset in (a) shows a schematic of an AM nanoribbon. The channels correspond to spin-dependent electron–electron and hole–hole transmission processes (equal-spin and spin-flip), shown for different Néel vector angles: (a)-(c) $\theta = 0$ and (d)-(f) $\theta = \pi/2$, respectively.}
\end{figure}

\begin{figure}[htbp]
\includegraphics[width=0.48\textwidth]{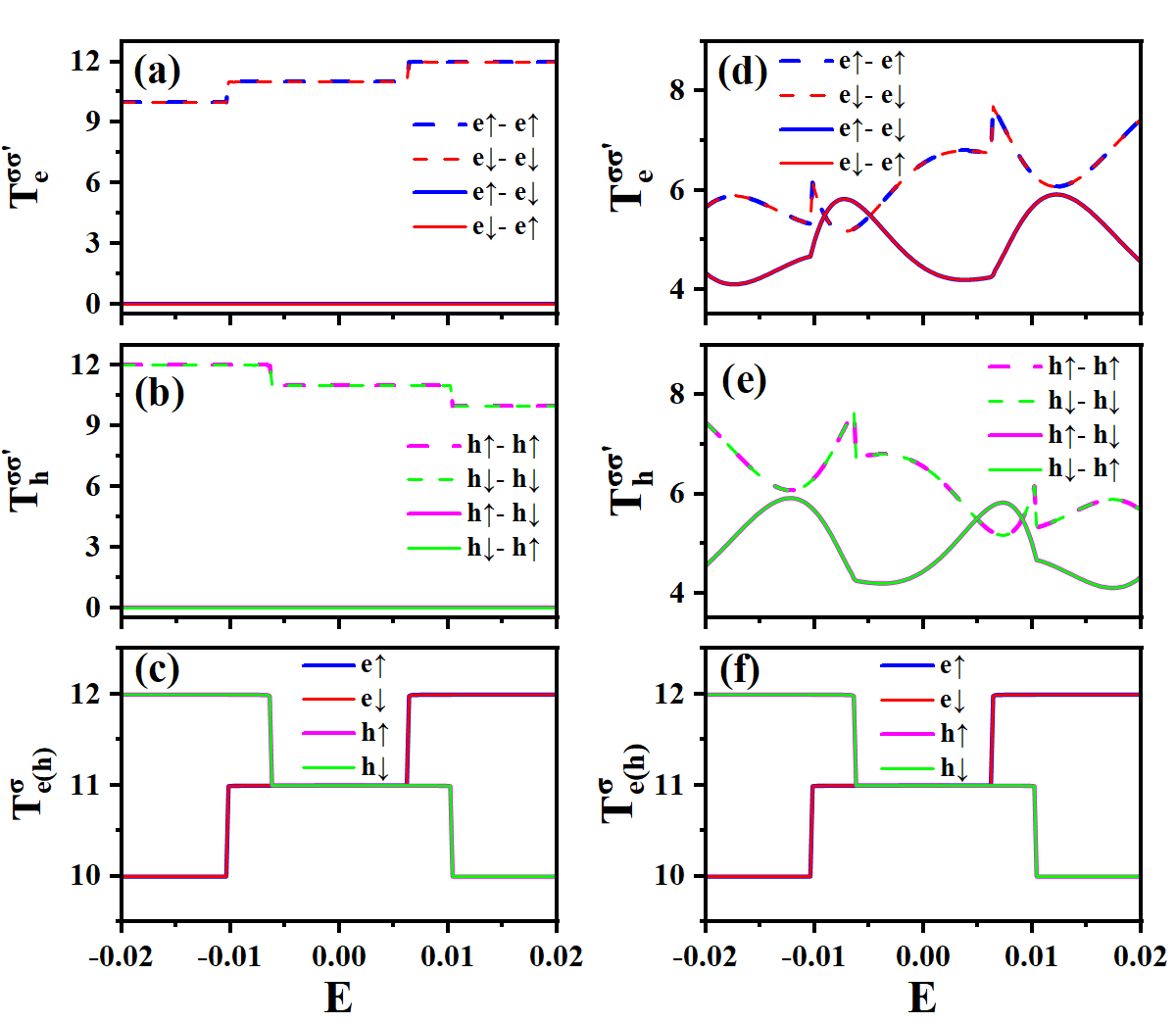}
\caption{\label{fig3} Spin-dependent transmission channels for electrons and holes in the AM nanoribbon at a fixed orientation angle $\alpha = \pi/4$, shown for different Néel vector angles: (a)-(c) $\theta = 0$ and (d)-(f) $\theta = \pi/2$. The channels include electron–electron and hole–hole transmission processes with both equal-spin and spin-flip components.}
\end{figure}

\begin{figure}[htbp]
\includegraphics[width=0.48\textwidth]{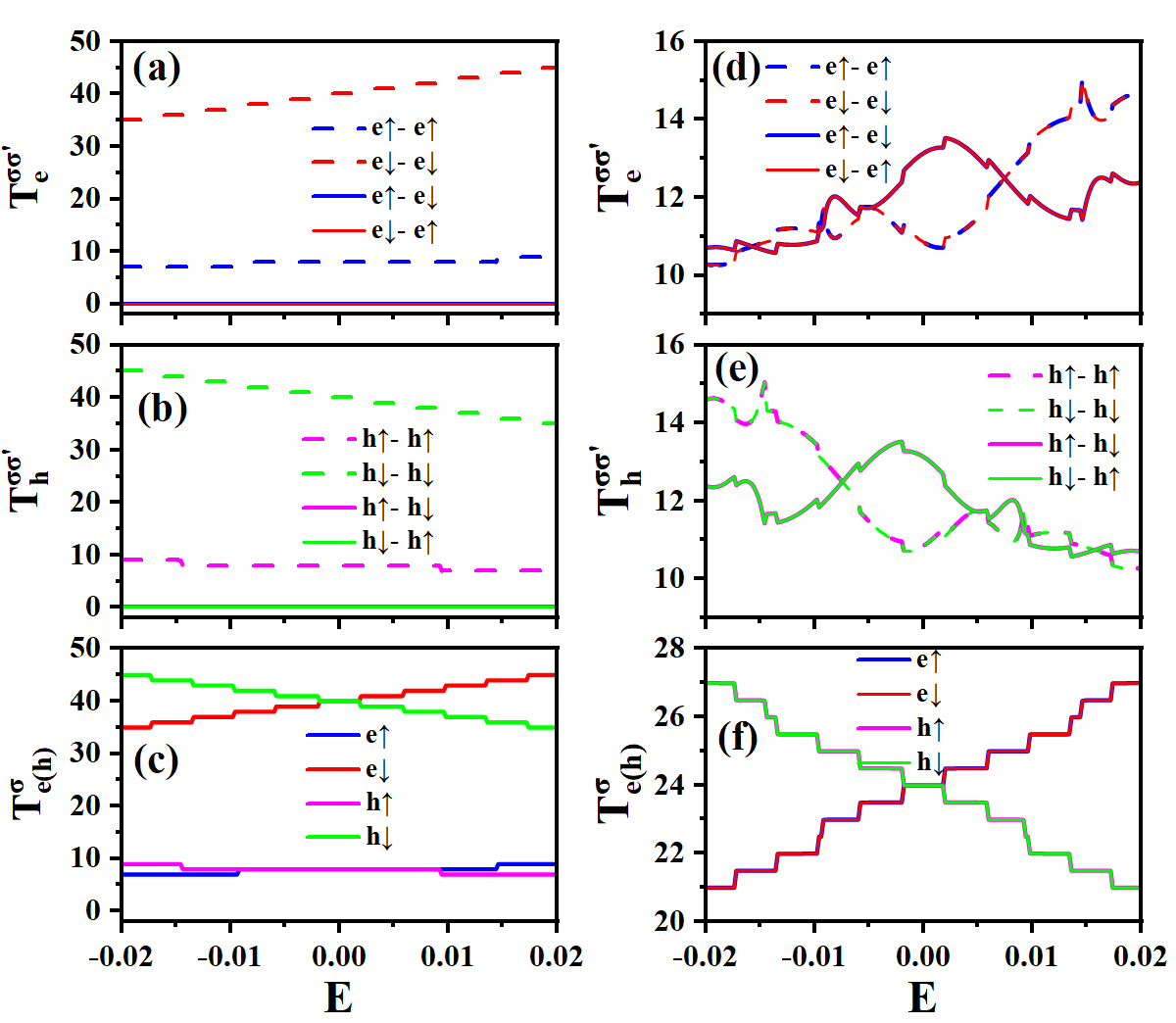}
\caption{\label{fig4} Transmission channels in the AM nanoribbon at a fixed orientation angle $\alpha = 0$. The channels characterize the spin character of electron–electron and hole–hole transmission, including both equal-spin and spin-flip processes. Results are shown for different Néel vector angles: (a)-(c) $\theta = 0$ and (d)-(f) $\theta = \pi/2$, respectively.}
\end{figure}

\begin{figure}[htbp]
\includegraphics[width=0.48\textwidth]{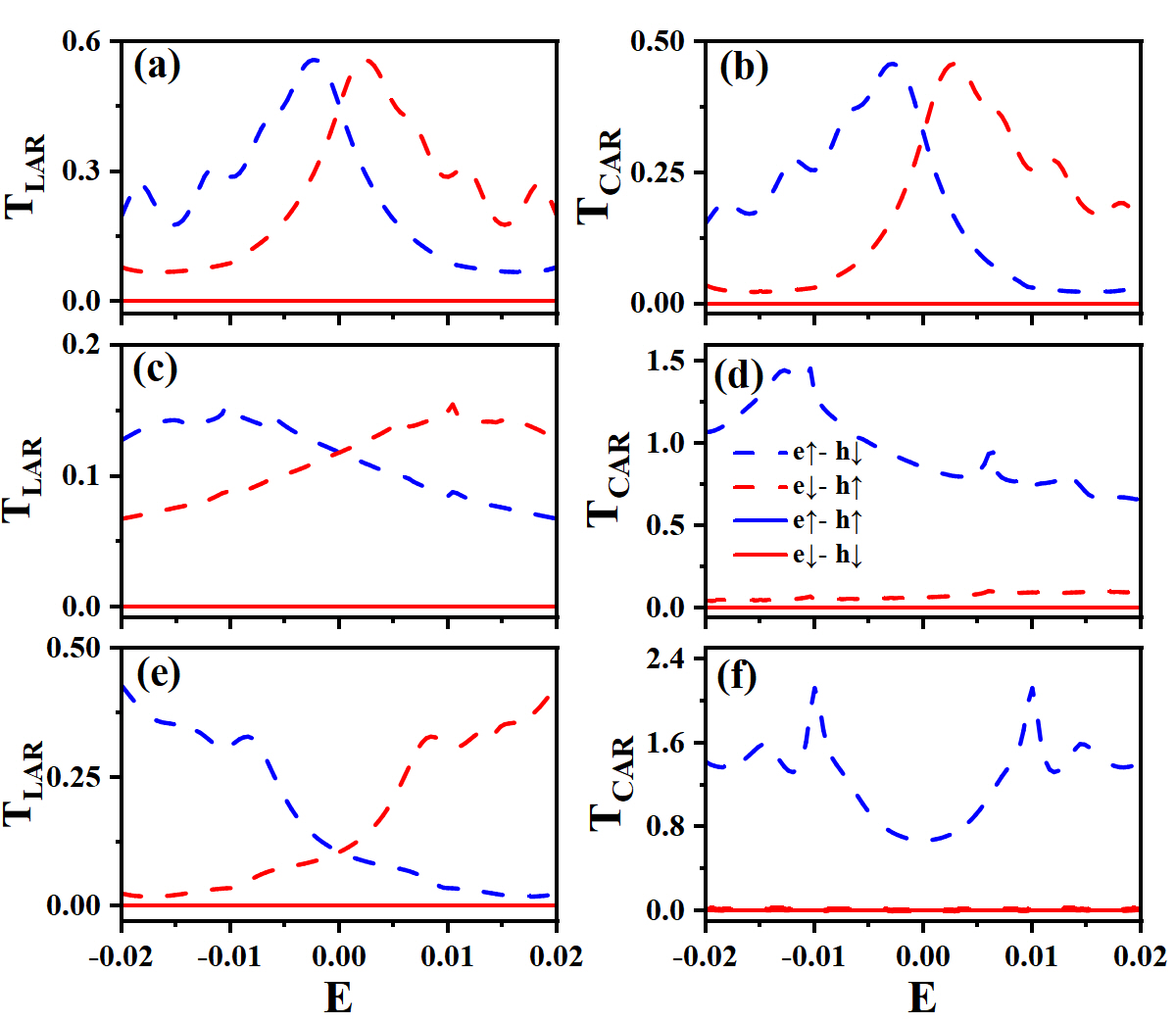}
\caption{\label{fig5} (a), (b) display the $T_{\mathrm{LAR}}$ and $T_{\mathrm{CAR}}$ for $\alpha_{L} = 0$ and $\alpha_{R} = 0$. (c), (d) correspond to $\alpha_{L} = 0$ and $\alpha_{R} = \pi/4$, while (e), (f) show $\alpha_{L} = 0$ and $\alpha_{R} = \pi/2$. Dashed (solid) lines indicate conventional (equal-spin) AR. The Néel vector angles are fixed at $\theta_{L} = 0$ and $\theta_{R} = 0$.}
\end{figure}

\section{RESULTS AND ANALYSIS}\label{SC3}

In this section, we study the dependence of local and crossed Andreev reflection in AM/SC/AM junctions on the crystal orientation angle $\alpha_{L(R)}$ and the Néel vector angle $\theta_{L(R)}$. Throughout the numerical calculations, the hopping energy is set to $t = 1$ and the lattice constant to $a = 1$. The altermagnetic strength and chemical potential are chosen as $t_J = 0.9$ and $\mu = 0.1$, respectively. The central superconducting region has a length $L = 80a$ and a width $W = 20a$.

\subsection{Spin-dependent transmission channels in altermagnets}

We first examine the effects of the crystal orientation angle $\alpha$ and the Néel vector angle $\theta$ on the spin-dependent transmission channels of electrons and holes in the AM nanoribbon, as shown in the inset of Fig.~\ref{fig2}(a). Figures~\ref{fig2}(a)–\ref{fig2}(c) display the transmission channels at $\theta = 0$, where $T_{e}^{\sigma\sigma'}$ and $T_{h}^{\sigma\sigma'}$ denote the transmission channels of electrons and holes, respectively. $T_{e(h)}^{\sigma}$ represents the total transmission associated with incident electrons (holes) of spin $\sigma$, defined as $T_{e(h)}^{\sigma} = \sum\limits_{\sigma'} T_{e(h)}^{\sigma \sigma'}$. Here, $\sigma$ and $\sigma'$ are the spins of the incident and transmitted particles, respectively, with $\sigma, \sigma' = \uparrow, \downarrow$. As shown in Figs.~\ref{fig2}(a) and~\ref{fig2}(b), only the dashed lines corresponding to equal-spin transmission are present, whereas the solid lines associated with spin-flip transmission vanish. This indicates that, in the absence of Néel vector rotation, neither electrons nor holes undergo spin flipping during transmission through the AM. The number of spin-up transmission channels for both electrons and holes is significantly larger than that of spin-down channels, which is consistent with the band structure shown in Fig.~\ref{fig1}(b). In Fig.~\ref{fig2}(c), because $\theta = 0$, spin-flip processes are absent and $T_{e(h)}^{\sigma }=T_{e(h)}^{\sigma \sigma }$.

With the orientation angle $\alpha = 0$, the Néel vector is rotated to $\theta = \pi/2$, the transmission channels are shown in Figs.~\ref{fig2}(d) -~\ref{fig2}(f). In Figs.~\ref{fig2}(d) and~\ref{fig2}(e), spin-flip transmission channels emerge, while the equal-spin transmission channels are simultaneously suppressed. This demonstrates that rotation of the Néel vector induces spin-flip processes for both electrons and holes. By comparing Figs.~\ref{fig2}(c) and~\ref{fig2}(f), we find that at $\theta = \pi/2$ the transmission channels for opposite spin species become identical. This comparison indicates that the spin-up and spin-down bands of electrons and holes are strongly mixed due to the rotation of the Néel vector.

Then, we change the orientation angle of the AM to $\alpha = \pi/4$, and the corresponding electron and hole transmission channels are shown in Fig.~\ref{fig3}. Figures~\ref{fig3}(a) -~\ref{fig3}(c) correspond to the case without Néel vector rotation. In this set, no spin-flip transmission channels are present. In contrast to the case of $\alpha = 0$, the equal-spin transmission channels for spin-up and spin-down completely overlap at $\alpha = \pi/4$, consistent with the band structure shown in Fig. 1(c).

Upon rotating the Néel vector to $\theta = \pi/2$, as shown in Figs.~\ref{fig3}(d) -~\ref{fig3}(f), spin-flip transmission channels again emerge, similar to those observed in Figs.~\ref{fig2}(d) and~\ref{fig2}(e). This further confirms that Néel vector rotation induces spin-flip transmission for both electrons and holes. Comparison between Figs.~\ref{fig3}(c) and~\ref{fig3}(f), it reveals that the transmission channels remain the same, in contrast to the phenomenon observed in Figs.~\ref{fig2}(c) and~\ref{fig2}(f). As inferred from the band structure in Fig.~\ref{fig1}(c), when the AM orientation angle $\alpha = \pi/4$, the spin-up and spin-down bands fully overlap, indicating identical numbers of spin-up and spin-down transmission channels. Rotation of the Néel vector then induces spin-flip processes between the two spin species without altering the total number of transmission channels.

Figure~\ref{fig4} shows the electron and hole transmission channels for $\alpha = \pi/2$. By comparing Figs.~\ref{fig1}(b) and~\ref{fig1}(d), the spin-up and spin-down energy bands are exchanged. Consequently, when $\alpha = \pi/2$, the exchange of the spin-up and spin-down energy bands is directly reflected in the transmission channels, leading to an interchange of the corresponding spin-dependent electron and hole transmission channels in Figs.~\ref{fig4}(a)–~\ref{fig4}(c). After rotating the Néel vector $\theta = \pi/2$, spin-flip occurs. As a result, the transmission channels shown in Figs.~\ref{fig4}(d)–\ref{fig4}(f) coincide with those in Figs.~\ref{fig2}(d)–\ref{fig2}(f).

\begin{figure}[htbp]
\includegraphics[width=0.48\textwidth]{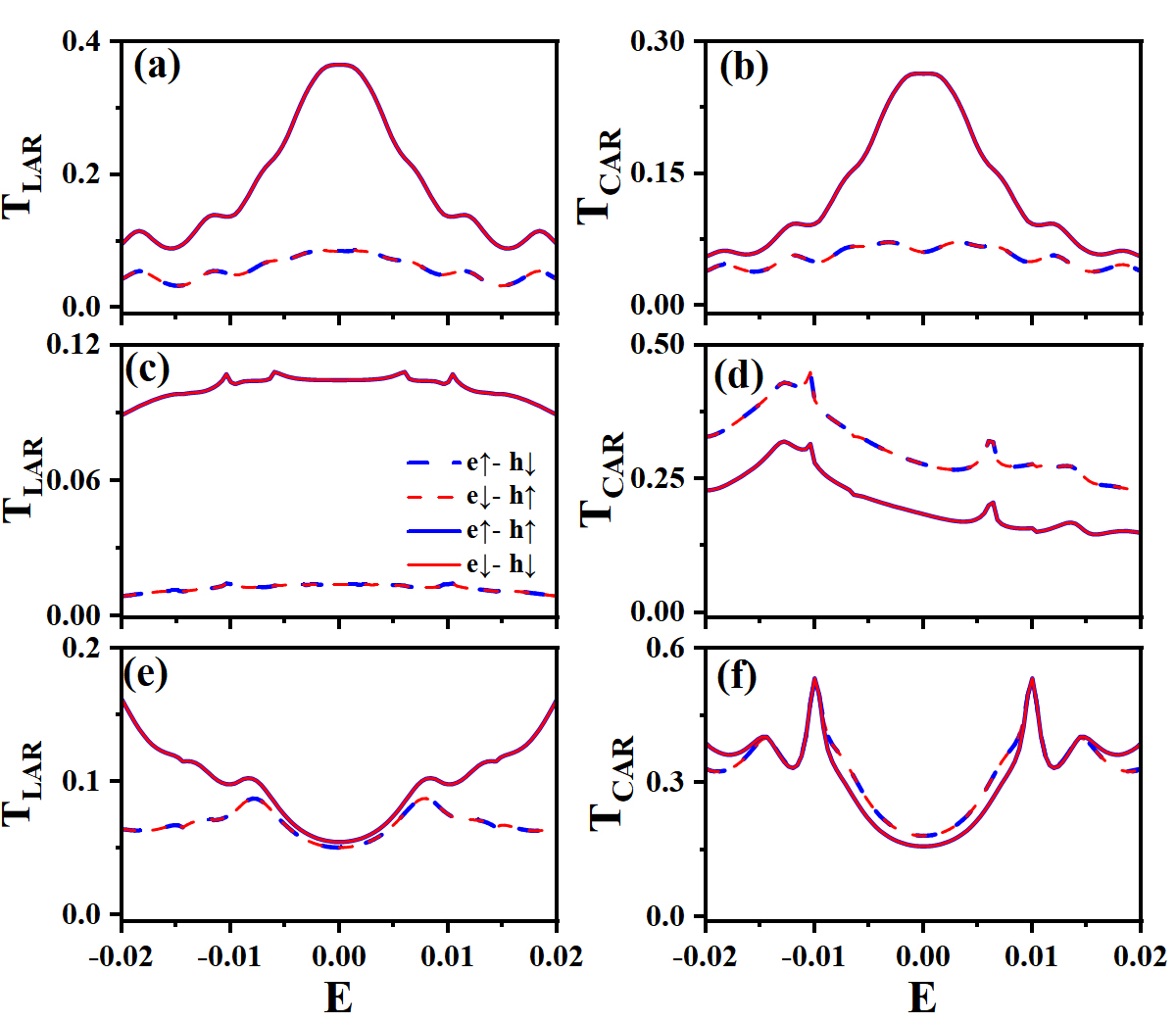}
\caption{\label{fig6} The $T_{\mathrm{LAR}}$ and $T_{\mathrm{CAR}}$ are plotted for different AM orientation angles. (a), (b) correspond to $\alpha_{L} = 0$ and $\alpha_{R} = 0$, (c), (d) for $\alpha_{L} = 0$ and $\alpha_{R} = \pi/4$, with (e), (f) corresponding to $\alpha_{L} = 0$ and $\alpha_{R} = \pi/2$. Dashed and solid lines represent conventional and equal-spin AR, respectively. The Néel vector angles are set to $\theta_{L} = \pi/2$ and $\theta_{R} = \pi/2$.}
\end{figure}

\subsection{Crystal Orientation and Néel Vector Effects on Andreev Reflection}

Next, we investigate the influence of the crystal orientation angles $\alpha_{L(R)}$ and the Néel vector angles $\theta_{L(R)}$ on AR in the AM/SC/AM junction. 

Andreev reflection is an electron–hole conversion process occurring at the interface between a superconductor and a normal conductor, in which an incident electron is converted into a hole while a Cooper pair is simultaneously formed inside the superconductor. Due to the superconducting energy gap, single-particle quasiparticle transport is suppressed inside the superconducting region. Instead, electrons with opposite spins form Cooper pairs through superconducting correlations, enabling charge transport through the superconductor. Consequently, reflected or nonlocal transmitted hole signals can be detected in the adjacent AM leads.

\begin{figure}[htbp]
\includegraphics[width=0.48\textwidth]{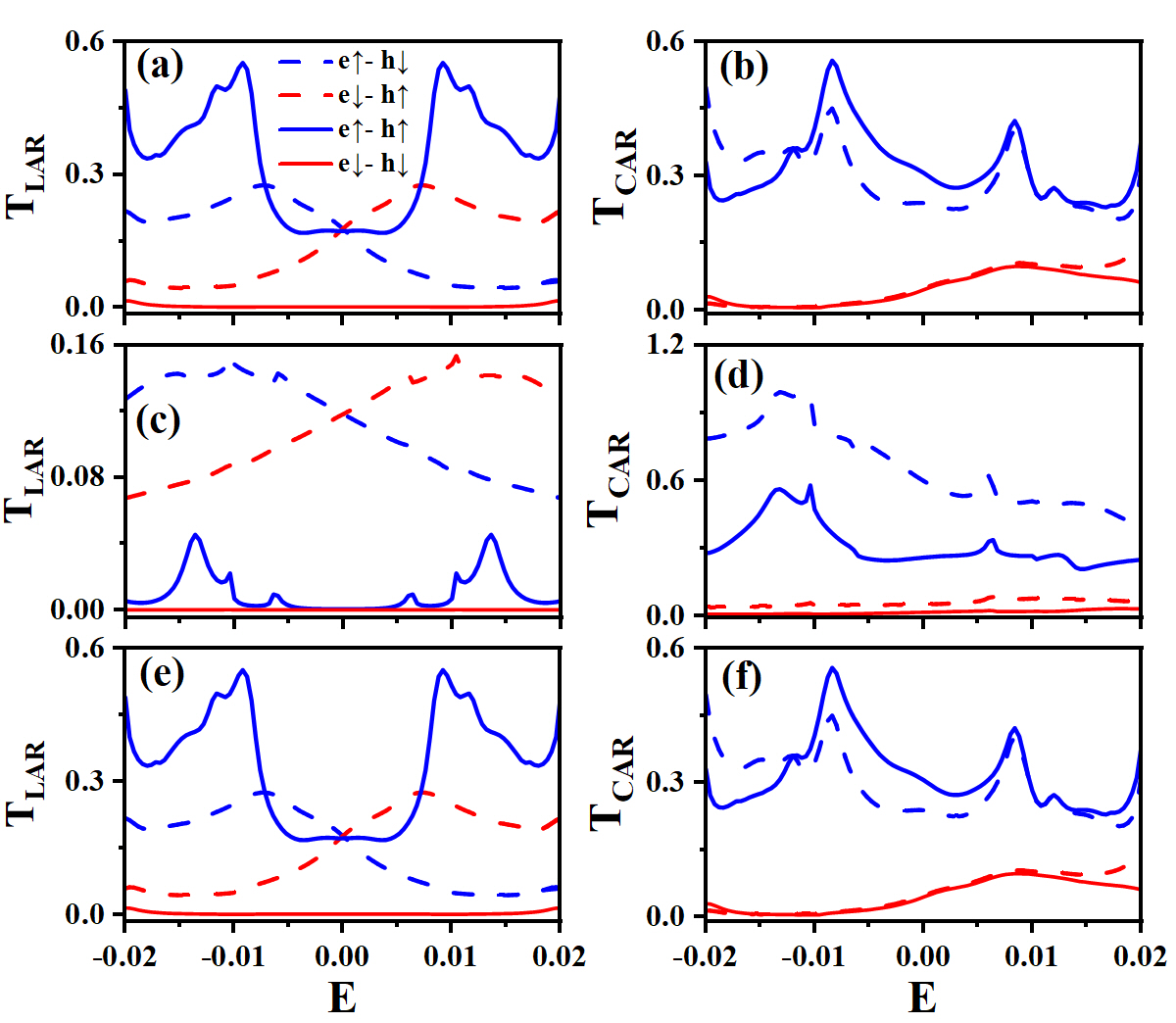}
\caption{\label{fig7} The $T_{\mathrm{LAR}}$ and $T_{\mathrm{CAR}}$ for AMs at different orientation angles. (a), (b) correspond to $\alpha_{L} = 0$ and $\alpha_{R} = 0$, (c), (d) for $\alpha_{L} = 0$ and $\alpha_{R} = \pi/4$, while (e), (f) for $\alpha_{L} = 0$ and $\alpha_{R} = \pi/2$. Dashed (solid) lines denote conventional (equal-spin) AR. The Néel vector angles are $\theta_{L} = 0$ and $\theta_{R} = \pi/2$ for all panels.}
\end{figure}

To better visualize the equal-spin local and nonlocal AR processes, in Figs.~\ref{fig5} -~\ref{fig9} all solid lines represent equal-spin local and nonlocal AR, whereas all dashed lines denote conventional local and nonlocal AR.

\subsubsection{Andreev Reflection with $\theta_{L} = \theta_{R} = 0$}

We first consider the case in which the Néel vectors remain unrotated, $\theta_{L} = \theta_{R} = 0$, as shown in Fig.~\ref{fig5}. The orientation angle of the left AM lead is fixed at $\alpha_L = 0$, while that of the right AM lead is varied as $\alpha_R = 0$, $\pi/4$, and $\pi/2$. The blue (red) dashed lines denote the conventional AR induced by incident spin-up (spin-down) electrons, whereas the blue (red) solid lines represent the equal-spin AR for incident spin-up (spin-down) electrons. Figures~\ref{fig5}(a),~\ref{fig5}(c), and~\ref{fig5}(e) show the local Andreev reflection coefficient $T_{\mathrm{LAR}}$ as a function of the incident energy $E$, while Figs.~\ref{fig5}(b),~\ref{fig5}(d), and~\ref{fig5}(f) display the crossed Andreev reflection coefficient $T_{\mathrm{CAR}}$.

When the Néel vectors are not rotated, the equal-spin AR channels (solid lines) vanish identically. This indicates that varying the crystal orientation angle alone does not induce spin-flip, and consequently no equal-spin AR occurs in this situation.

Figures~\ref{fig5}(a) and~\ref{fig5}(b) display $T_{\mathrm{LAR}}$ and $T_{\mathrm{CAR}}$ for $\alpha_L = 0$ and $\alpha_R = 0$, respectively. Since the left and right AM leads have the same crystal orientation angle, the conventional AR for both spin-up and spin-down incident electrons are symmetric with respect to $E = 0$, which is consistent with the band structure shown in Fig.~\ref{fig1}(b). When the orientation angle of the right lead $\alpha_R = \pi/4$, the CAR shown in Fig.~\ref{fig5}(d) becomes asymmetric with respect to $E = 0$. Specifically, the conventional $T_{\mathrm{CAR}}$ for incident spin-up electrons is enhanced, while that for incident spin-down electrons is suppressed. By comparing the transmission channels presented in Figs.~\ref{fig2}(c) and~\ref{fig3}(c), we find that near $E = 0$ and for $\alpha_R = \pi/4$, the spin-up hole transmission channels are significantly reduced, whereas the spin-down hole channels are increased relative to the $\alpha_R = 0$ case. This redistribution of hole transmission channels leads to an enhancement of the conventional CAR for incident spin-up electrons.

In addition, CAR becomes asymmetric with respect to $E = 0$ when $\alpha_R=\pi/4$. As shown in Fig.~\ref{fig1}(c), the available hole transmission channels are also asymmetric with respect to $E = 0$, with more hole channels available below the $E = 0$ than above it. Since the CAR process involves the conversion of an incident electron into a transmitted hole in the opposite lead, the asymmetry of the hole transmission channels further enhances CAR in the $E < 0$ region, resulting in the asymmetric behavior observed in Fig.~\ref{fig5}(d).

When the orientation angle $\alpha_R = \pi/2$, the conventional CAR induced by incident spin-up electrons is further enhanced and becomes symmetric with respect to $E = 0$ in Fig.~\ref{fig5}(f). This is because the number of spin-down hole transmission channels reaches its maximum at $\alpha_R = \pi/2$. By comparing the band structures in Figs.~\ref{fig1}(b) and~\ref{fig1}(d), the spin-up electron band at $\alpha_R = 0$ and the spin-down hole band at $\alpha_R = \pi/2$ are symmetric with respect to $E = 0$. The dominant contribution to the conventional CAR for incident spin-up electrons is strengthened. This result is consistent with our previous conclusion: for $\alpha_R=\pi/2$, the numbers of transmission channels for spin-up and spin-down electrons and holes are reversed compared with the case of $\alpha=0$.

\subsubsection{Andreev Reflection with $\theta_{L} = \theta_{R} = \pi/2$}

Figure~\ref{fig6} shows the $T_{\mathrm{LAR}}$ and $T_{\mathrm{CAR}}$ as functions of the incident energy $E$, with the Néel vector $\theta_{L}=\theta_{R}=\pi/2$. As shown in Figs.~\ref{fig6}(a),~\ref{fig6}(c), and~\ref{fig6}(e), when $\theta_{L(R)} = \pi/2$, the equal-spin LAR remains the dominant contribution, independent of the orientation angle $\alpha_R$ of the right lead. In Figs.~\ref{fig2}(d) and~\ref{fig2}(e), the numbers of spin-flip channels for both electrons and holes increase significantly at $\theta=\pi/2$. Electrons and holes propagating through either lead can undergo spin-flip, which enhances the probability of generating equal-spin AR. From the overlap of electron and hole transmission channels shown in Fig.~\ref{fig2}(f), it can be inferred that $T_{\mathrm{LAR}}$ first increases and then decreases with energy increasing, which is consistent with Fig.~\ref{fig6}(a). Since Figs.~\ref{fig6}(a) and~\ref{fig6}(b) correspond to identical orientation angles of the left and right AM leads ($\alpha_L=\alpha_R$), the band structures of the two leads are the same. Consequently, $T_{\mathrm{CAR}}$ exhibits a similar nonmonotonic dependence on energy, increasing first and then decreasing, as shown in Fig.~\ref{fig6}(b).

In contrast, the equal-spin CAR is not dominant throughout the range $\alpha_R \in [0, \pi/2]$ and exhibits a pronounced dependence on $\alpha_R$, as illustrated in Figs.~\ref{fig6}(b), ~\ref{fig6}(d), and ~\ref{fig6}(f). When $\alpha_R=\pi/4$, Fig.~\ref{fig3}(e) shows that the total number of spin-flip hole transmission channels is smaller than that of the equal-spin channels. As a result, the equal-spin contribution to $T_{\mathrm{CAR}}$ in Fig.~\ref{fig6}(d) is weaker than the conventional $T_{\mathrm{CAR}}$. When $\alpha_R=\pi/2$, a comparison between Figs.~\ref{fig2} and~\ref{fig4} indicates that at $\theta=\pi/2$, the spin-dependent transmission channels become identical for the two orientation angles $\alpha_R=0$ and $\alpha_R=\pi/2$. Comparing Figs.~\ref{fig6}(a) and~\ref{fig6}(b) with Figs.~\ref{fig6}(e) and~\ref{fig6}(f) reveals that, despite the transmission channels being the same for $\alpha_R = 0$ and $\alpha_R = \pi/2$, the resulting $T_{\mathrm{LAR}}$ and $T_{\mathrm{CAR}}$ differ markedly. This difference arises because the Néel vector and the crystal orientation angle act on fundamentally different Hilbert spaces in the AM system. The Néel vector corresponds to rotations in spin space, whereas the crystal orientation angle represents a rotation in real space. As a reault, even when the available transport channels are equivalent, the Andreev reflection remains highly sensitive to the crystal orientation of the AM.

Upon rotating the Néel vector, equal-spin AR emerges, and the equal-spin AR for incident spin-up and spin-down electrons coincide. The conventional AR for opposite spin species also overlap. This behavior is consistent with our earlier conclusion drawn from Figs.~\ref{fig2} –~\ref{fig4}.

\begin{figure}[htbp]
\includegraphics[width=0.48\textwidth]{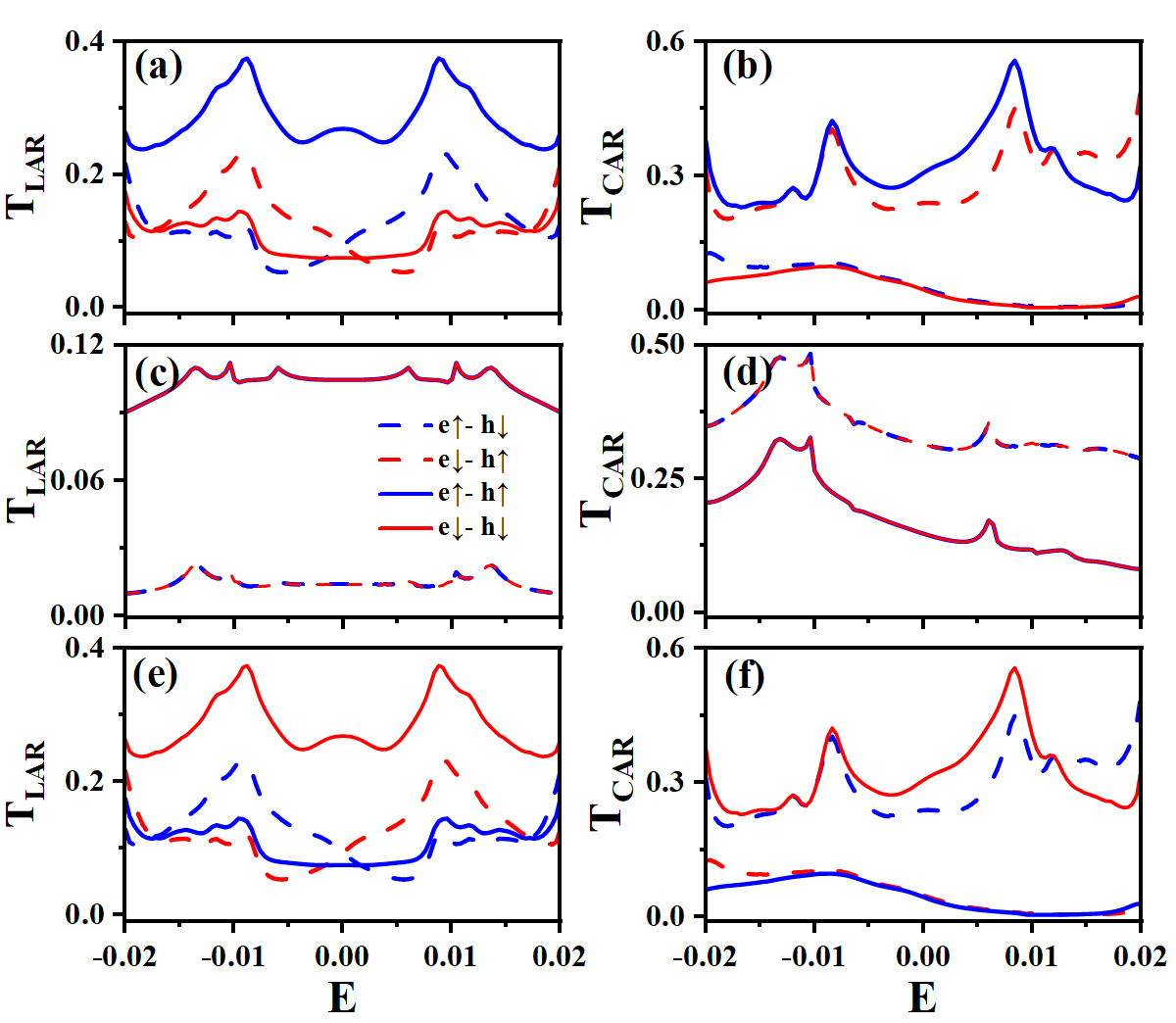}
\caption{\label{fig8} Illustrated here are the $T_{\mathrm{LAR}}$ and $T_{\mathrm{CAR}}$ for various AM orientation angles. (a), (b) show $\alpha_{L} = 0$ and $\alpha_{R} = 0$, (c), (d) $\alpha_{L} = 0$ and $\alpha_{R} = \pi/4$, (e), (f) $\alpha_{L} = 0$ and $\alpha_{R} = \pi/2$. Dashed lines correspond to conventional AR, whereas solid lines indicate equal-spin processes. The Néel vector angles are fixed at $\theta_{L} = \pi/2$ and $\theta_{R} = 0$.}
\end{figure}

\subsubsection{Andreev Reflection with $\theta_{L} = 0$, $\theta_{R} = \pi/2$}

We further investigate the AR processes in the AM/SC/AM junction for $\theta_L=0$ and $\theta_R=\pi/2$, as shown in Fig.~\ref{fig7}. From Figs.~\ref{fig7}(a),~\ref{fig7}(c), and~\ref{fig7}(e), it is evident that the Néel vector in the left lead is not rotated, equal-spin LAR still occurs. This phenomenon originates from spin-flip at the SC/AM interface: a fraction of incident electrons experience spin flipping at the interface, and some of these spin-flipped electrons subsequently form Cooper pairs with electrons of opposite spin within the superconducting region, giving rise to equal-spin LAR. In contrast, equal-spin CAR arises from two distinct contributions. One contribution comes from holes generated by Cooper pairs formed before spin-flip, which subsequently undergo spin flipping in the right lead. The other contribution originates from holes produced by Cooper pairs formed after spin-flip of electrons. As a result, even when only the Néel vector in the right AM lead is rotated, both equal-spin LAR and equal-spin CAR can coexist in the system. Because the number of spin-up electron transmission channels is largest for $\alpha_L = 0$, the equal-spin AR and the conventional CAR induced by incident spin-up electrons provide the dominant contributions.

In Figs.~\ref{fig7}(c) and~\ref{fig7}(d), corresponding to $\alpha_L = 0$ and $\alpha_R = \pi/4$, the equal-spin AR is suppressed due to the limited number of spin-flip transmission channels following the rotation of the Néel vector, as illustrated in Figs.~\ref{fig3}(d) and~\ref{fig3}(e). By comparing Figs.~\ref{fig7}(a),~\ref{fig7}(b),~\ref{fig7}(e), and~\ref{fig7}(f), the results for $\alpha_R = 0$ and $\alpha_R = \pi/2$ are identical. This equivalence can be understood from the fact that changing $\alpha_R$ from 0 to $\pi/2$ merely interchanges the spin-up and spin-down bands. Due to the rotation of the Néel vector, spin-band mixing occurs. As demonstrated in Figs.~\ref{fig2} and~\ref{fig4}, under these conditions the channels for $\alpha_R = 0$ and $\alpha_R = \pi/2$ become identical, resulting in the same AR.

\subsubsection{Andreev Reflection with $\theta_{L} = \pi/2$, $\theta_{R} = 0$}

Finally, we consider the case with Néel vector $\theta_L = \pi/2$ and $\theta_R = 0$, the results are shown in Fig.~\ref{fig8}. When $\theta_L = \pi/2$, a portion of the incident electrons undergoes spin-flip in the left AM lead and subsequently enters the superconducting region, where they form Cooper pairs with electrons of opposite spin. This process gives rise to equal-spin LAR and equal-spin CAR. Meanwhile, electrons that do not experience spin flipping propagate into the superconducting region and form Cooper pairs, generating conventional LAR and CAR. Some holes generated by conventional LAR may undergo spin-flip in the left lead, contributing to equal-spin LAR. As a result, both equal-spin and conventional AR processes coexist even though only the Néel vector in the left lead is rotated.

Figures~\ref{fig8}(a) and~\ref{fig8}(b) present the AR processes at $\alpha_R = 0$, with dominant contributions from equal-spin LAR and CAR of incident spin-up electrons and conventional CAR of incident spin-down electrons. In Fig.~\ref{fig2}(c), the number of spin-down electron and hole channels is smaller than that of spin-up channels, resulting in strong reflection of spin-down electrons at the SC/AM interface and enhanced Cooper-pair formation in the superconducting region, which gives rise to the AR observed in Figs.~\ref{fig8}(a) and~\ref{fig8}(b).
At $\alpha = \pi/4$, the numbers of spin-up and spin-down transmission channels for both electrons and holes are equal, as indicated by Fig.~\ref{fig3}(c). Equal-spin AR contributions from incident spin-up and spin-down electrons become identical, and conventional AR for the two spin species also coincide, as seen in Figs.~\ref{fig8}(c) and~\ref{fig8}(d). 
When $\alpha = \pi/2$, the spin-up electron and hole transmission channels are smaller than the spin-down channels, as shown in Fig.~\ref{fig4}(c). At $\alpha_R = \pi/2$, strong reflection of incident spin-up electrons at the SC/AM interface promotes Cooper-pair formation in the superconducting region, resulting in equal-spin LAR and equal-spin CAR dominated by incident spin-up electrons, whereas conventional CAR is dominated by incident spin-down electrons, as shown in Figs.~\ref{fig8}(e) and~\ref{fig8}(f).

By comparing the results for $\alpha_R=\pi/4$ in Figs.~\ref{fig7} and~\ref{fig8}, the distinct roles of the left and right AM leads in spin-flip processes become evident. When $\theta_L=0$ and $\theta_R=\pi/2$, the spin-flip of electrons and holes occurs predominantly in the right lead, so the CAR is mainly governed by electrons incident with a single spin orientation, achieve a larger CAR. In contrast, when $\theta_L=\pi/2$ and $\theta_R=0$, the spin-flip takes place primarily in the left lead. Due to the strong mixing of spin channels in the left lead, incident spin-up and spin-down electrons contribute equally to the AR processes, and the equal-spin LAR becomes the dominant contribution.

\begin{figure}[htbp]
\includegraphics[width=0.48\textwidth]{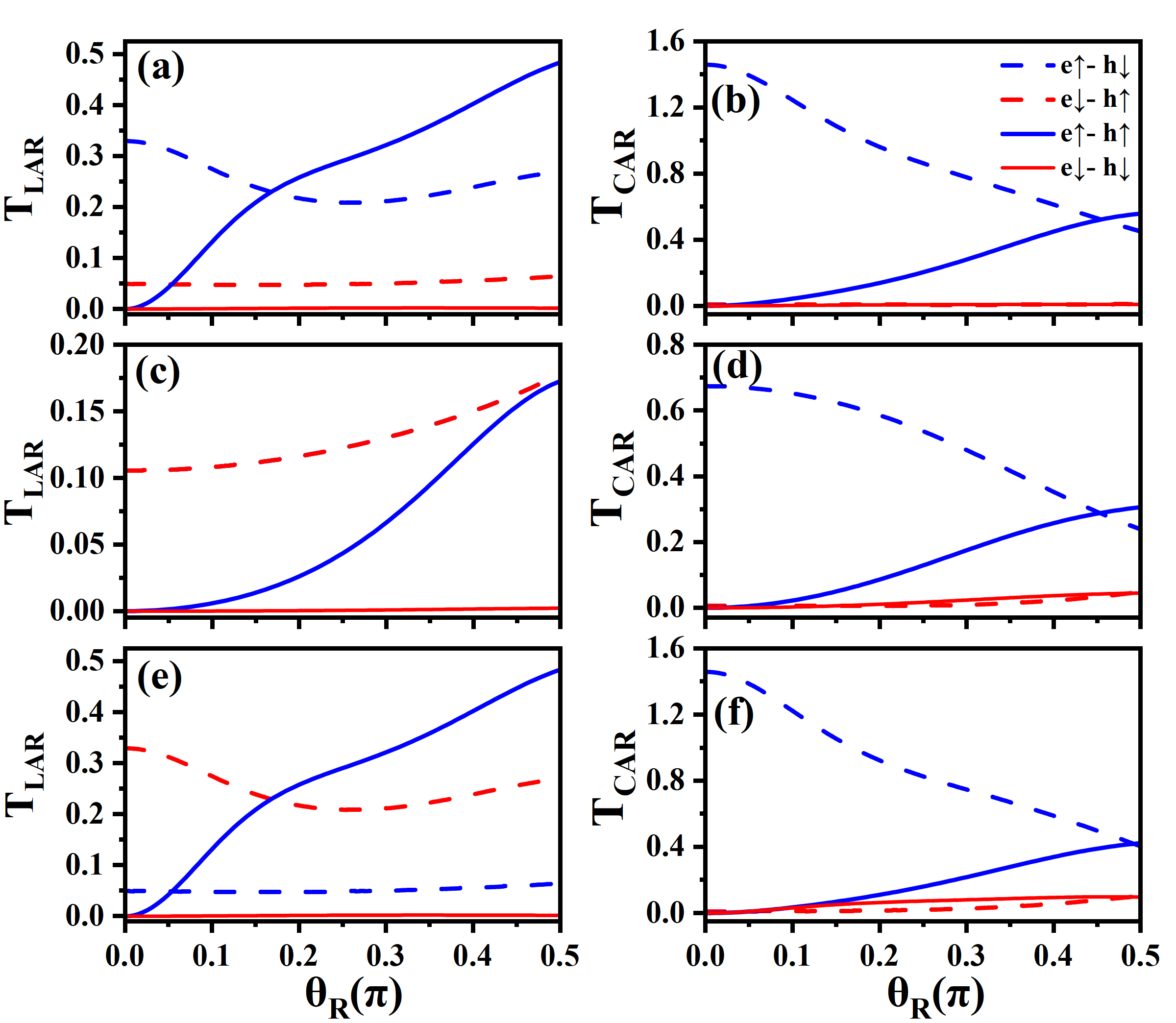}
\caption{\label{fig9} Dependence of $T_{\mathrm{LAR}}$ and $T_{\mathrm{CAR}}$ on the Néel vector angle in the right AM lead at different incident energies. (a), (b) $E = -0.0084$, (c), (d) $E = 0$, (e), (f) $E = 0.0084$. The dashed (solid) curves represent conventional (equal-spin) AR. The crystal orientation angles are set to $\alpha_{L} = 0$ and $\alpha_{R} = \pi/2$, while the Néel vector angle in the left lead is fixed at $\theta_{L} = 0$.}
\end{figure}

To further investigate the influence of the Néel vector on equal-spin AR, we compare the results shown in Figs.~\ref{fig5}–\ref{fig8} for different crystal orientation angles. We find that relatively strong equal-spin AR can be achieved for the configurations $\alpha_L = 0$, $\alpha_R = 0$ and $\alpha_L = 0$, $\alpha_R = \pi/2$ when the Néel vector satisfy $\theta_L = 0$ and $\theta_R = \pi/2$. Based on these results, we further study the dependence of different AR on the Néel vector $\theta_R$ for the case $\alpha_L = 0$, $\alpha_R = \pi/2$, and $\theta_L = 0$ at representative energies $E = -0.0084$, $0$, and $0.0084$, as shown in Fig.~\ref{fig9}.

Figures~\ref{fig9}(a), (c), and (e) show the variations of $T_{LAR}$ with $\theta_R$ at different energies. For $E = \pm 0.0084$, when $\theta_R$ becomes sufficiently large, the equal-spin LAR induced by incident spin-up electrons exceeds the conventional LAR contribution. At $E = 0$, the equal-spin LAR also exhibits an increasing tendency with increasing $\theta_R$. Figures~\ref{fig9}(b), (d), and (f) present the corresponding $T_{CAR}$ as a function of $\theta_R$. When $\theta_R$ is small, the conventional CAR dominates the transport process. As $\theta_R$ approaches $\pi/2$, the equal-spin CAR becomes comparable to the conventional CAR. These results indicate that both local and nonlocal equal-spin AR are significantly enhanced by increasing the Néel vector $\theta$. Physically, the rotation of the Néel vector strengthens spin mixing in the altermagnetism band structure, thereby promoting equal-spin AR in the AM/SC/AM junction.

These results demonstrate that the combined effects of the crystal orientation angle and the Néel vector give rise to distinct orientation-dependent AR in the AM/SC/AM junction. In particular, the Néel vector plays a crucial role in the emergence of equal-spin AR in AM/SC/AM junctions.

\section{CONCLUSIONS}\label{SC4}

In summary, We investigated the Andreev reflection phenomenon in an altermagnet/superconductor/altermagnet (AM/SC/AM) junction, focusing on the roles of the crystal orientation angle $\alpha_{L(R)}$ and the Néel vector $\theta_{L(R)}$.
Without Néel vector rotation, no equal-spin Andreev reflection appears in the system. Tuning the orientation angles of the AM leads gives rise to orientation-dependent conventional CAR. Under specific conditions, conventional CAR is mainly dominated by incident spin-up electrons, while the contribution from spin-down electrons becomes negligible.
Once the Néel vector is rotated in either lead, spin-flip is activated, giving rise to both equal-spin local and crossed Andreev reflection. Depending on the specific combination of crystal orientation and Néel vector, distinct AR phenomenon dominated by different spin species can be realized. These results identify the rotation of the Néel vector as the direct origin of equal-spin AR in AM/SC/AM junctions.
Our study demonstrates that the interplay between crystal orientation and Néel vector rotation enables orientation-dependent Andreev reflection processes.

\begin{acknowledgments}
This work was supported by the National Natural Science Foundation of China (Grant No. 11874139) and the Natural Science Foundation of Hebei Province (Grant No. A2019205190).
\end{acknowledgments}

\section*{DATA AVAILABILITY}
The data that support the findings of this study are available from the corresponding author upon reasonable request.

\nocite{*}

\bibliography{apssamp}

\end{document}